\documentclass{llncs}

\usepackage[normalem]{ulem}

\begin{document}

\title{Revealing digital documents}
\subtitle{Concealed structures in data%
\footnote{$[$v1$]$ 2011-05-29. To be presented at
the TPDL 2011 Doctoral Consortium, Berlin.}}


\author{Jakob Vo\ss}
\institute{Verbundzentrale des GBV (VZG), G\"ottingen, Germany,\\
\email{jakob.voss@gbv.de}}

\maketitle

\begin{abstract}
This short paper gives an introduction to a research project to analyze
how digital documents are structured and described. Using a phenomenological
approach, this research will reveal common patterns that are used in data,
independent from the particular technology in which the data is available.
The ability to identify these patterns, on different levels of description, is 
important for several applications in digital libraries. A better understanding
of data structuring will not only help to better capture singular 
characteristics of data by metadata, but will also recover intended structures 
of digital objects, beyond long term preservation.
\keywords{data, data description, metadata, data modeling, patterns}
\end{abstract}


\section{Introduction}


Given the growing importance of digital documents in libraries, the theoretical
underpinning of data in library and information science is still insufficient.
The majority of bibliographic descriptions only exist in digital form. 
Increasingly documents only exist as digital objects, which impacts on traditional
concepts such as `document`, `page`, `edition`, and `copy`.  Meanwhile most 
metadata consists of digital documents that describe other digital documents. 
With the advent of networked environments,\footnote{An ongoing trend that is
most visible in hypes, for instance, today, cloud computing and Semantic Web.} 
these documents basically exist as streams of bits, abstracted from any 
storage medium and location. Although in practice concrete forms, such as 
`files`, `records', and `objects', are dealt with, these forms are rather 
different views on the same thing, than inherent properties of a digital document. 
So what is this `same thing' if you talk about a digital document? It has been
shown that the nature of documents can better be defined in terms of function
rather than format \cite{Buckland1998}, and the key properties, which 
constitute and identify a document, depend on context \cite{Yeo2010}. This
highlights the importance of descriptive metadata to put data in context,
but it does not eliminate the need to actually look at data at some level
of description. In practice, we often have to deal with heterogeneous documents
provided as data that must be indexed and preserved, or with metadata, that is 
aggregated from diverse sources, without exact description of the data on a 
higher level.

This paper proposes that a deeper look at data is required, to reveal how digital 
documents are actually structured and described. The question should not be 
answered by simply pointing to concrete technologies and formats, which are 
subject to rapid change and obsolescence, but at a more fundamental level. 
The main hypothesis of this research is that all methods to structure and 
describe data share common patterns, independent from technology and level
of description.


\section{Background and related works}


The concept of data is used in many disciplines with various meanings. 
Ballsun-Stanton explores how different individuals understand data in different
``philosophies of data'' \cite{BallsunStanton2010}: the concept can range from 
the product of objective, reproducible measurements (``data as hard numbers'') 
to the product of any recorded observations (``data as observations''), or
processable encodings of information and knowledge (``data as bits''). With this
research I commit to the third philosophy, which is often found in computer 
science and in library and information science.\footnote{This does not imply 
that results will not be applicable to other philosophies of data. Revealed data 
patterns may also reflect typical structures of data observation and 
measurements. However, this is beyond the scope of this work.} However, both 
disciplines do not use data as a core concept but relate it to information 
as the main topic of interest.
The growing amount of freely available ``open data'' and tools to analyze 
this data has brought up ideas of ``data science'' and ``data journalism''.
Both deal with aggregating, filtering, and visualizing large sets of data, 
based on statistical methods of data analysis. The growth of data-driven 
science combined with principles of Open Access also raises the awareness 
of the need to publish and share data sets. Library institutions begin to
recognize this need and start to provide infrastructure for collecting
and identifying research data \cite{DLib2011}. Data discussed in this 
context is mainly seen as ``data as observations'' and the main concern of
data science evangelists seems to be ``big data'', that is ``when the size 
of the data itself becomes part of the problem'' \cite{Loukides2010}.
The problem, that I want to tackle, does not depend on the size of the data 
or on problems of preservation \cite{Rosenthal2010}, but on the inherent 
complexity of data, independent from its applications. 

While disciplines
that deal with physical documents, such as codicology and palaeography, have 
long been acknowledged as part of library and information science, there is 
no established curriculum of data studies as yet.
The best examination of data by libraries so far can be found in long-term
preservation of digital material and in metadata research and practice. 
The former is still in an early stage of development.\footnote{By definition
you can only speak retrospectively of successful long-term preservation, but 
most digital objects are too young to judge.}
It provides two general strategies to  
cope with the rapid change and decay of technologies: either you need to 
emulate the environment of digital objects or you must regularly migrate 
them to other environments and formats. Both strategies require good
descriptions of the data to be archived. When time passes, these descriptions
themselves become subject of preservation and digital objects may get buried in
nested layers of metadata.
%
%
Metadata research deals with data on 
a more explicit level. Although metadata has become one of the core concepts of
library and information science, there is no commonly agreed upon definition. 
The general ``data about data'' definition at least makes clear that metadata is 
data about something. Coyle's definition of metadata as something constructed, 
constructive, and actionable \cite{Coyle2010} highlights the relevance of
function and context as Buckland \cite{Buckland1998} and Yeo \cite{Yeo2010} 
do for documents. As a result there are numerous ways to describe the 
same object by data and the same data can describe different things. Metadata
research provides at least some guidelines for interoperability by metadata 
registries, application profiles, and crosswalks. However, in practice a lot of
manual work is needed to make use of metadata, because context and function 
are not fully known or creators of data just do not comply to assumed standards.
Currently, the Resource Description Framework (RDF) and persistent identifiers
promise to solve most problems. However, as also confirmed by the preliminary 
results below, there is no silver bullet in data description. Data is always 
a simplified, context-dependent image of the information, knowledge, or reality 
where an attempt has been to encode it in data. Some good criticism of the 
expressive power of particular data encoding languages has been given by Kent
\cite{Kent1978,Kent1988,Kent2003}.


Patterns as structured methods of describing good design practice, were first
introduced by Alexander et al. in the field of architecture \cite{Alexander1977}.
In their words ``each pattern describes a problem which occurs over and over
again in our environment, and then describes the core of the solution to that
problem.'' Patterns were later adopted in other fields of engineering, 
especially in (object-oriented) software design \cite{Beck1987,Gamma1994}.
There are some works that describe pattern for specific data structuring or 
data modeling languages, among them Linked Data in RDF \cite{Dodds2010}, 
Markup languages \cite{Dattolo2007a}, data models in enterprises%
\cite{Hay1995,Silverston2001}, and meta models \cite{Hay2006,Silverston2009}.
A general limitation of existing approaches is the focus to one specific 
formalization method. This practical limitation blocks the view to 
more general data patterns, independent from a particular encoding, and it
conceals blind spots and weaknesses of a chosen formalism. 

\section{Method}

A preliminary analysis of different structuring methods shows that each 
data language highlights some structuring features that are then overused 
and even conceal intended structures of data. For instance, the nesting and 
order of elements in an XML document can be chosen with intent. However, they
can also be chosen in an arbitrary manner because an XML document must always
be an ordered tree. For this reason we cannot rely only on official 
descriptions and specifications to reveal patterns in data. Most 
existing approaches to analyze data structuring are either normative 
(theoretical descriptions how data should be), or empirical but limited. 
Existing empirical approaches only view data at one level of description,
in order to have a base for statistical methods (data mining) and other automatic
methods (machine learning). In contrast, I use a phenomenological research 
method that includes all aspects of data structuring and description. Beside 
technical standards that specify data, software that shapes data, typical 
examples of data, and at how data is actually used by people, must also
be considered.

The phenomenological method views data as social artifacts, that cannot be
described from an absolute, objective point of view. Instead data are studied
as ```phenomena`: appearances of things, or things as they appear in our 
experience'' \cite{Smith2009}. The analysis begins with a detailed review of
methods and systems for structuring and describing data, from simple character
encodings to data languages and even graphical notations. The focus is on 
conceptual properties, while details of implementation, such as performance 
and security, are only mentioned where they show how and why specific 
techniques have evolved. 

\section{Preliminary results}
The first outcome of this work is a broad typology of existing methods to 
structure and describe data. These methods are normally described as data 
codes, systems, languages, or models without consistent terminology jointly
among technologies. The following groups of methods can be identified, each
with a primary but not exclusive purpose:

\begin{itemize}
\item \textbf{character and number encodings} to express data
\item \textbf{identifiers} and \textbf{query languages} to identify data
\item \textbf{file systems} and \textbf{databases} to store data
\item \textbf{data structuring languages} and \textbf{markup languages} 
      to structure data
\item \textbf{schema languages} to define and constrain data
\item \textbf{conceptual modeling languages} to abstract and describe data
\end{itemize}

These methods are rarely discussed together as general structures with data 
as their common domain. Instead a strong focus on trends and families of basic
technologies is found, that often concentrate on one specification or implementation.
Examples include; the dominance of the Structured Query Language (SQL) and the
hype around the extensible markup language (XML) in the late 1990s. With size
and speed as a main driving force of development, there is little progress at the
conceptual level. An example of this being the large gap between
research and practice in conceptual modeling languages, which are mostly used
in form of an oversimplification of the Entity-Relationship Model (ERM)
\cite{Simsion2007}. 

The main empirical part of the analysis consists of a detailed description and
placement of the most relevant instances and subgroups from the typology above.
It is shown how each structuring method has its strengths and limitations, and 
how each method shapes digital objects independent from the object's 
characteristic properties. A deeper look at data also shows that the most
influential technologies of data structuring are not used in one exact and 
established form, but they occur as groups of slightly differing variants.
For example, the set of data expressible in RDF/XML differs from the 
full RDF triple model. This triple model with URIs, blank nodes and literals, 
also has different characteristics and limitations depending on which 
technology (serialization, triple store, SPARQL, reasoners etc.) and 
which entailment regime (simple, RDF, OWL etc.) is applied. Other examples 
include; SQL databases, which substantially differ from the original 
relational database model, and the family of XML-related standards. In 
many cases confusion originates from differences between syntaxes and
implementations on one side and abstract models on the other.

To some degree, common patterns can be derived from specific systems
by modeling them in a higher level modeling language, such as ERM
and Object Role Modeling (ORM), or in schema and ontology languages 
such as Backus Naur Form (BNF), XML Schema (XSD) and the Web Ontology 
language (OWL). In software engineering it is common practice 
to use domain specific modeling languages in nested layers of abstraction
\cite{Kelly2008}. These languages exist in many variants as tools to 
communicate between levels of description. As a result, each language 
highlights a specific subset of patterns and makes other patterns less 
visible or more difficult to apply. Typical instances of data further show
that in practice, patterns and levels of abstraction often overlap and that 
methods of structuring are often used against their original purpose. Typical 
examples include; the creation of dummy values for non-existing mandatory 
elements and the use of separators to add lists to non-repeatable fields. 
It appears that in practice it is often difficult to judge which properties 
of data are intended and which arise as artifacts from the constraints of a 
given modeling language. The example of XML was already mentioned above: XML
structures data in form of an ordered tree, but many instances of XML documents
use this feature to apply other patterns but hierarchy and strict ordering.
Figure~\ref{fig:sequencepattern} shows an example from a yet to be finished
catalog of data patterns.

\begin{figure}
\centering
\fbox{
  \begin{minipage}{\linewidth}
\begin{description}
\item[name]
 \hfill \\
 \uline{sequence} pattern

\item[idea] 
 \hfill \\
  strictly order multiple objects, one after another

\item[context]
 \hfill \\
  a \uline{collection} of multiple objects

\item[implementations] \hfill
\\ $\bullet$  
  If objects have a \uline{known size}, they can be directly concatenated.
  If objects have \uline{same size}, this results in the \uline{array} pattern.
\\ $\bullet$
  The \uline{separator} pattern can be used to separate each object from its
  successor object. To distinguish objects and separators, this implies the
  \uline{forbidden objects} pattern. If separators may occur directly after 
  each other, this may also imply the \uline{empty object} pattern.
\\ $\bullet$
  You can link one object to its successor with an \uline{identifier}.
  To avoid link structures that result in other patterns
  (\uline{tree}, \uline{graph}, \ldots) additional constraints
  must apply.
\\ $\bullet$ If objects have consecutive \uline{positions}, a sequence is 
  implied by their order.
\item[examples] \hfill 
\\ $\bullet$
  string of ASCII characters (\uline{array})
\\ $\bullet$
  string of Unicode characters in UTF-8 (each character has \uline{known size})
\\ $\bullet$
  `\texttt{Kernighan and Ritchie}' (sequence with `\texttt{ and }' as separator)
\\ $\bullet$
  $extract \rightarrow transform$, 
  $transform \rightarrow load$, (sequence of linked steps)

\item[counter examples] \hfill \\
  files in a file system, records in a database table, 
  any unordered collection

\item[motivation] \hfill \\
  sequences are a natural method to model one-dimensional phenomena, for
  instance sequences of events in time. As digital storage is structured
  as sequence of bits, sequences seem to be the natural form of data and
  counterexamples, such as formal diagrams and visual programming languages,
  are often not considered as data.

\item[problems] \hfill \\
  empty sequences and sequences of only one element are difficult to spot,
  like in other \uline{collection} patterns.

\item[similar patterns] \hfill \\
  without context, sequences are difficult to distinguish from other 
  \uline{collection} patterns. Many implementations of other patterns
  use sequences on a lower level.

\item[implied patterns]\hfill \\
  \uline{position} pattern

\item[specialized patterns]\hfill \\
  \uline{array}, \uline{ordered set}, \uline{ring}
\end{description}
\end{minipage} } 
\caption{Example of a pattern description. Pattern names are underlined.}
\label{fig:sequencepattern}
\end{figure}

\section{Evaluation and application}


The preliminary results show a large variety of methods to structure and 
describe data. The research hypotheses can be confirmed, as common patterns 
like identifiers, repeatability, grouping, sequences and ordering are 
used on all levels in different variants and explicitly. The ability to 
identify and apply these patterns is crucial for several applications in 
digital libraries. Some patterns are already recognized, but the results 
show that it lacks a more systematic view, independent from the constraints 
of particular technologies. A better understanding of methods to structure 
and describe data can help both, the creation of data and its consumption.
These applications are shortly illustrated in the following.

Creation of data in libraries is most notably present as creation of 
metadata. This process is guided by complex cataloging rules and 
specialized formats. Both are deeply intertwined and often criticized
as barriers to innovation. However, simpler forms of metadata do not provide 
a solution \cite{Tennant2004}. Remarkably, alternatives are most visible
as technologies, for instance XML \cite{Tennant2002} or RDF \cite{Coyle2010}.
Despite the strengths of each technology, it is unlikely that one method
will provide the ultimate tool to express all metadata. Instead a look
at metadata pattern can aid the construction of more precise and 
interoperable metadata that better captures an object's unique 
characteristics. The nature of patterns in general shows that data 
creation is no an automatic process, but a creative act of design. 
Recognizing the artificial nature of data will to some degree free 
data designers from apologies and unquestioned habits that are justified
as enforced by natural needs or technical requirements.

Consumption of data can benefit even more from an understanding of data 
patterns. Since the invention of digital computers, technologies and 
formats rapidly change. The fluctuation will unlikely slow down because 
it is also driven by trends, as progress in data description (in contrast
to quantitative data processing) is difficult to measure. The results show 
that many description methods result in other structures than originally 
intended, when the the patterns that are actually applied are examined. 
Relevant structures are less visible, if you concentrate on single technologies.
Knowledge of general data patterns can therefore help to reveal concealed
structures in digital documents. This application could be named ``data
archeology''.\footnote{The most related existing discipline is digital 
forensics. It has a more specific scope and its application to more 
complex and heterogeneous methods of data structuring, e.g. databases,
is in an early stage of development \cite{Olivier2009}.}
Data archeology, in contrast to long-term preservation, which
tries to prevent the need of the former, deals with the retrospective analysis
of incompletely defined or unknown data. Similar to traditional archeology,
data archeology belongs to the humanities, as it involves study of the 
cultural context of data creation and usage. Existing techniques from
computer science, like data mining and knowledge discovery, provide
useful tools to discover detailed views on data. However, they cannot reveal 
its meaning as part of social practice. Data patterns can provide a 
contribution to intellectual data analysis, which is needed to underpin 
and interpret algorithmic data analysis.

Beside the creation of a catalog of the most common data patterns as
basic primitives and derived patterns, there are some open tasks that
may be answered by the analysis described in this paper. It is assumed
that no closed system or meta-system can fully describe all aspects of
practical data. This thesis could be proved at least for formal systems
of description based on results of G\"odel \cite{Goedel1931}. Further
research, which will probably not be covered fully in this work, includes 
how to best find known patterns in given data using semi-automatic methods
and which methods are best suited to express a given set of patterns.

In any case, libraries can benefit from a general understanding of data and
data patterns, at least as deep as the current understanding of physical 
publication types and material.



\bibliographystyle{splncs03}
\bibliography{phdvoss_tpdl2011_v1} 

\end{document}